\documentclass[cameraready]{Interspeech}
\title{OnDA: On-device Channel Pruning for Efficient Personalized Keyword Spotting}

\author[affiliation={1}, correspondingauthor]{Matteo}{Risso}
\author[affiliation={1}]{Alessio}{Burrello}
\author[affiliation={1}]{Daniele}{Jahier Pagliari}
\address{
  $^1$ Politecnico di Torino, Italy
}

\email{matteo.risso@polito.it, alessio.burrello@polito.it, daniele.jahier@polito.it}


\usepackage{xcolor}
\usepackage{comment}
\usepackage{amsmath,amssymb}
\usepackage{graphicx}
\usepackage{booktabs}
\usepackage{multirow}
\usepackage{url}
\usepackage{siunitx}

\usepackage{makecell}

\usepackage{algorithm}
\usepackage{algpseudocode}

\begin{document}
\maketitle
\begin{abstract}
  Always-on keyword spotting (KWS) demands \emph{on-device adaptation} to cope with user- and environment-specific distribution shifts under tight latency and energy budgets.
  This paper proposes, for the first time, coupling \textit{weight} adaptation (i.e., on-device training) with \emph{architectural} adaptation, in the form of online structured channel pruning, for personalized on-device KWS.
  Starting from a state-of-the-art self-learning personalized KWS pipeline, we compare data-agnostic and data-aware pruning criteria applied on in-field pseudo-labelled user data. On the HeySnips and HeySnapdragon datasets, we achieve up to 9.63$\times$ model-size compression with respect to unpruned baselines at iso-task performance, measured as the accuracy at 0.5 false alarms per hour.
  When deploying our adaptation pipeline on a Jetson Orin Nano embedded GPU, we achieve up to 1.52$\times$/1.57$\times$ and 1.64$\times$/1.77$\times$ latency and energy-consumption improvements during online training/inference compared to weights-only adaptation.
\end{abstract}

\section{Introduction}
Keyword spotting (KWS) represents a key building block for always-on voice interfaces, enabling wake-word and command detection under continuous audio streams~\cite{lopez2021deep}.
In practical deployment scenarios, KWS systems face different challenges such as (i) dealing with keywords not seen during training~\cite{rusci2023customization}, (ii) large inter-speaker variability~\cite{rusci2024self}, (iii) domain shift from training to in-field acoustic conditions~\cite{cioflan2024device}, and (iv) stringent on-device constraints on memory, compute, and energy~\cite{rusci2024self, cioflan2024device}.
These challenges motivate the development of personalized and adaptive KWS methods that can improve in the target domain after deployment, without relying on large labeled datasets.

Recent work has studied on-device self-learning for personalized KWS on ultra-low-power sensors by combining pseudo-labeling with incremental gradient-based fine-tuning \cite{rusci2024self}.
However, we argue that the efficiency of personalized KWS can be significantly improved by adapting online not only the \textit{weights} of a Deep Neural Network (DNN), but also its \textit{architecture}.

Classically, DNN architecture optimizations (in the form of pruning, neural architecture search, etc.) are treated as offline, pre-deployment steps~\cite{li2016pruning,yu2022hessian}. Intuitively, however, the same distribution shifts that make on-device fine-tuning beneficial may also influence well-performing architectures. For instance, a more or less aggressively pruned model might be needed to reach a certain accuracy, depending on the target keyword, the speaker's voice, the acoustic conditions, etc.

% In particular, it is reasonable to expect that online model optimization (e.g., structured pruning) can be beneficial when performed alongside online weight updates, since both aim to tailor the model to the user/domain while respecting tight compute and energy budgets.
% At the same time, model compression is often treated as an offline, pre-deployment step 

%This paper explores such complementary perspective: structured pruning as an \emph{online} procedure that complements on-device adaptation, improving efficiency while preserving (or improving) task performance.
In this paper, we therefore investigate On-device Adaptation (OnDA) pipelines that employ structured channel pruning \emph{online} to complement weights adaptation.
We select structured pruning for its effectiveness and compatibility with training/inference infrastructure, but OnDA's core principles are extensible to other forms of compression.
% , improving efficiency while preserving (or improving) task performance.
%
% Our core idea is simple: during online operation, we prune unimportant DNN channels based on user-specific data to meet a target compression, \textit{before} or \textit{while} fine-tuning weights.

% We compare data-agnostic global L1-norm channel pruning \cite{li2016pruning} with data-aware Hessian-Aware Pruning (HAP) scores computed via Hutchinson trace estimation on in-field user data \cite{yu2022hessian}.
%Beyond choosing the pruning criterion, we study \textbf{when} pruning should be applied: we explore both offline and online pruning, including combinations of offline+online pruning under both data-agnostic and data-aware scoring. Finally, we consider a pipeline that applies data-agnostic pruning \textbf{after} online adaptation, so that the updated weight distribution better reflects the target domain before compression.
Through experiments with data-agnostic and data-aware pruning methods applied at different stages of the adaptation pipeline, we demonstrate that optimizing the architecture using in-distribution data is generally superior to offline alternatives. This confirms the above intuition, which is non-trivial given the trade-offs between data quality and quantity. In fact, while the OnDA approach uses data that more closely match the in-field distribution, it consequently has access to \textit{significantly fewer samples} for post-pruning fine-tuning, which also relies on \textit{pseudo-labels} rather than hard ground truths.
We summarize our main contributions as follows:
\begin{itemize}
  \item We formulate and evaluate multiple edge-deployment pipelines that combine offline pretraining, structured pruning (offline/online, data-agnostic/data-aware), and on-device self-learning adaptation.
  \item We report Pareto frontiers for accuracy with False Alarm Rate ($\text{FAR}_h$) $=0.5$ false alarms/hour versus model size on the HeySnips and HeySnapdragon datasets, using the ResNet15 and DS-CNN-L from~\cite{rusci2024self} as baseline architectures.  We achieve up to 3.33$\times$ and 9.63$\times$ model compression at iso-task performance with those baselines, while also showing that pruning with domain data is more effective than directly fine-tuning a smaller (offline-pruned) network.
  \item We provide NVIDIA Jetson Orin Nano latency and energy measurements, achieving up to 1.57$\times$/1.93$\times$ lower inference latency, and 1.77$\times$/2.07$\times$ lower inference energy-consumption at iso-task performance compared to the baseline DNNs of~\cite{rusci2024self} on GPU and CPU, respectively. Analyzing these results, we further conclude that data-aware pruning offers superior advantages over data-agnostic alternatives, as it can be applied \textit{at the start} of the adaptation phase, thereby reducing the cost of subsequent fine-tuning (and inference).
  % , which demonstrates that OnDA pipelines are stronger alternatives to common offline-only compression approaches.
\end{itemize}
%The OnDA code is released as open source at: \texttt{omitted-for-double-blind-review}.

%
\section{Background and Related Work}
%
%\subsection{On-Device Learning and Personalization for KWS}
\textbf{On-Device Learning and Personalization for KWS:}
On-device personalization aims to adapt a keyword detector to a target user and/or acoustic environment, often with only a few labelled samples.
ProtoNet formulations \cite{snell2017prototypical} (detailed in Sec.~\ref{sec:baseline_selflearning}) are common in few-shot and open-set KWS \cite{yang2022personalized, parnami2022few, rusci2023customization}.
The work of \cite{rusci2024self} proposes a self-learning pipeline for user-specific personalization in an open-set scenario where the personalized keyword belongs to a previously unseen class (i.e., not observed during pretraining). A pretrained ProtoNet architecture is employed to pseudo-label incoming utterances based on embedding similarity with a few user-provided samples. Finally, the pseudo-labelled dataset is used to perform on-device gradient-based fine-tuning.
An orthogonal direction is to adapt the model to the acoustic domain rather than to a specific user/keyword. For instance, \cite{cioflan2024device} studied on-device domain learning for KWS under changing ambient noise conditions on low-power extreme-edge systems.

\looseness=-1
\textbf{Pruning and Model Optimization for Edge KWS:}
Efficiency for KWS has been pursued through compact architectures (e.g., CNN/DS-CNN/ResNet variants) \cite{sainath2015cnn,tang2017resnet,coucke2019dilated,zhang2017hello}, quantization \cite{mishchenko2019lowbit}, pruning \cite{wang2022keyword}, and hardware-aware neural architecture search \cite{risso2022lightweight,busia2022nas}.
Hessian-based pruning has shown strong results across a range of tasks~\cite{yu2022hessian}, yet remains under-explored for KWS.
In particular, Hessian-based criteria are a promising direction, especially in low-data regimes where data-efficient importance estimates are desirable \cite{yu2022hessian}. Hessian-Aware Pruning (HAP) is a representative approach that proposes channel-wise pruning based on a Hessian-trace-weighted magnitude score computed efficiently via Hutchinson's trace estimation \cite{yu2022hessian}.
Nonetheless, pruning and architectural optimization in general have been applied only in an offline (pre-deployment) setting.
In contrast, OnDA leverages \emph{in-field user data} to compute pruning scores and investigates pruning to reduce the cost of both on-device fine-tuning and inference in personalized KWS.
\section{Methods}
\begin{figure}[t]
  \centering
  \includegraphics[width=\linewidth]{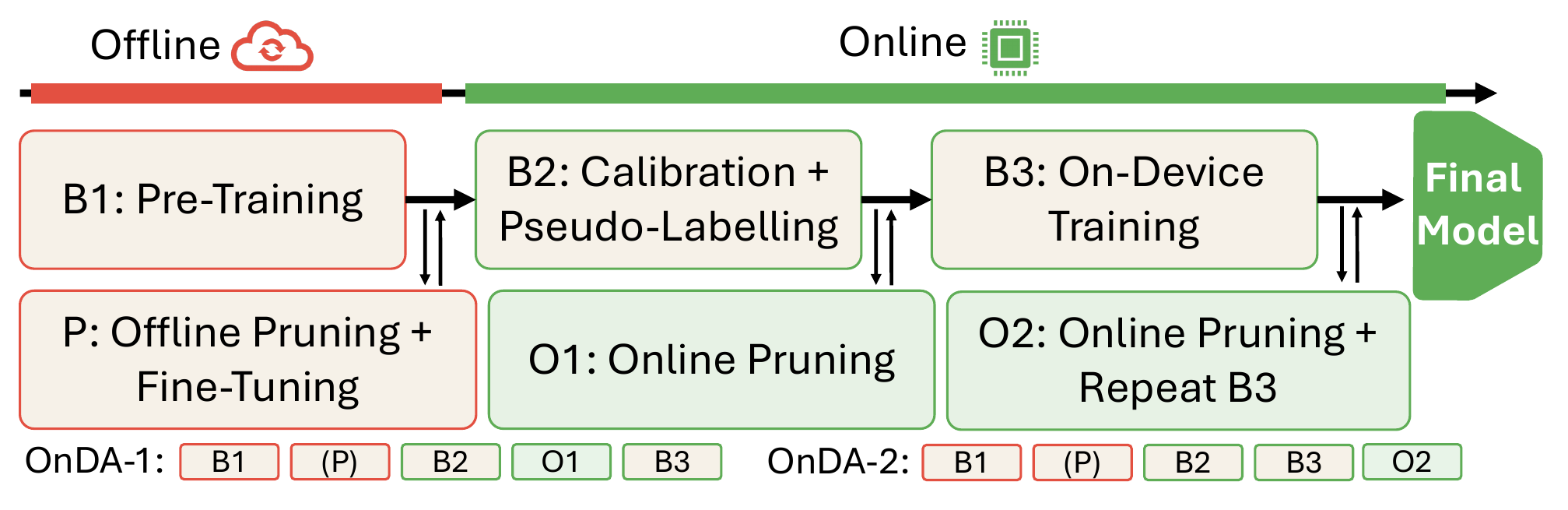}
  \vspace{-0.6cm}
  \caption{\textsc{Baseline} and \textsc{OnDA} pipelines steps.}
  \label{fig:method_overview}
\vspace{-0.5cm}
\end{figure}
The top of Fig.~\ref{fig:method_overview} (\textsc{B1}, \textsc{B2}, and \textsc{B3}) summarizes the self-learning KWS pipeline from~\cite{rusci2024self} that we take as baseline, which is summarized in Sec.~\ref{sec:baseline_selflearning}. On the bottom, \textsc{P} represents a ``classic'' offline pruning, whereas \textsc{O1} and \textsc{O2} are the additional online pruning steps proposed in our work. We describe OnDA pipelines in Sec.~\ref{sec:onda_pipelines}, and detail the pruning strategies that we select to realize them in Sec.~\ref{sec:pruning_strategies}.

\subsection{Baseline self-learning pipeline} \label{sec:baseline_selflearning}
We follow the self-learning protocol of \cite{rusci2024self}.
Let $f_{\theta}: \mathcal{X}\rightarrow \mathbb{R}^{d}$ denote a so-called ProtoNet DNN~\cite{snell2017prototypical}, mapping an input audio segment $x\in\mathcal{X}$ to an embedding $z=f_{\theta}(x)$.

The initial step of the pipeline (\textsc{B1}) deals with the offline pretraining of $f_{\theta}$.
Given a labeled multi-class and multi-speaker dataset $\mathcal{D}_{\mathrm{pre}}=\{(x_i,y_i)\}_{i=1}^{N}$ with labels $y_i\in\{1,\dots,K\}$, $f_{\theta}$ is pretrained using the triplet loss of Eq. \ref{eq:triplet}.
Each triplet is built by sampling from $\mathcal{D}_{\mathrm{pre}}$ two positive samples sharing the same class, and one negative from a different class, i.e., $(x_{p_1},x_{p_2},x_n)$ such that $y_{p_1}=y_{p_2}\neq y_n$.
Let $z_{p_1}=f_{\theta}(x_{p_1})$, $z_{p_2}=f_{\theta}(x_{p_2})$, and $z_n=f_{\theta}(x_n)$.
The loss for a single triplet is computed as:
\begin{equation}
  \mathcal{L}_{\mathrm{tri}}(\theta)=\max\Big(\lVert z_{p_1}-z_{p_2}\rVert_2^2-\lVert z_{p_1}-z_n\rVert_2^2+\alpha,\ 0\Big)
  \label{eq:triplet}
\end{equation}
where $\alpha>0$ is the margin hyper-parameter.

Then, the pretrained network $f_\theta$ is deployed on the target platform and the calibration and pseudo-labelling phase (\textsc{B2}) is executed.
During this phase, a user not present in the training set provides a small labelled set of $M$ positive examples of an unseen keyword (i.e., a class not present in $\mathcal{D}_{\mathrm{pre}}$), denoted as $\mathcal{E}=\{x^{+}_j\}_{j=1}^{M}$.
A keyword prototype is computed as:
\begin{equation}
  p = \frac{1}{M}\sum_{j=1}^{M} f_{\theta}(x^{+}_j).
  \label{eq:prototype}
\end{equation}

Next, incoming audio segments $x$ are pseudo-labelled by computing the Euclidean distance $d(x)=\lVert f_{\theta}(x)-p\rVert_2$ and comparing it with two thresholds $\gamma_{+}<\gamma_{-}$.
The segment is pesudo-labelled as positive if $d(x)\le \gamma_{+}$ and as negative if $d(x)\ge \gamma_{-}$; samples within $\gamma_{+}<d(x)<\gamma_{-}$ are discarded. The thresholds are calibrated following \cite{rusci2024self}.
Pseudo-labeled audio segments yield the set $\mathcal{D}_{\mathrm{ft}}=\{(x_t,\tilde{y}_t)\}$.

Finally, we perform on-device training (\textsc{B3}) by fine-tuning the embedding network $f_\theta$ using the same triplet-loss objective of Eq.~\eqref{eq:triplet}, forming triplets by sampling pseudo-labelled positives and negatives from $\mathcal{D}_{\mathrm{ft}}$. 
% In the rest of the manuscript we will refer to this pipeline as \textsc{Baseline}.

\begin{figure*}[!h]
  \centering
  \includegraphics[width=0.9\linewidth]{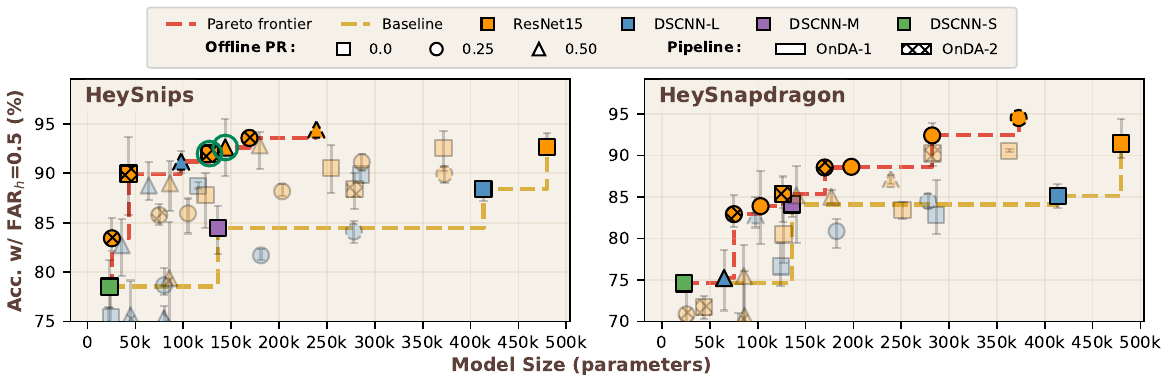}
  \vspace{-0.3cm}
  \caption{Pareto fronts for personalized KWS under different OnDA pipelines compared with baseline \cite{rusci2024self}.}
  \label{fig:pareto}
  \vspace{-0.4cm}
\end{figure*}

\subsection{OnDA pipelines} \label{sec:onda_pipelines}
% We propose two OnDA pipelines to improve personalized KWS efficiency beyond the \textsc{Baseline} of Sec.~\ref{sec:baseline_selflearning}.

Let $\mathcal{A}$ be the network architecture (i.e., a choice of channels per layer) for the embedding network $f_\theta$.
We denote by $\mathcal{A}' = \mathcal{P}_{\rho}^{(\cdot)}(\mathcal{A};\mathcal{B})$ a structured channel-pruning operator that transforms $\mathcal{A}$ into a pruned architecture $\mathcal{A}'$ at a pruning ratio $\rho$.
The operator may be \emph{data-aware} and use a batch $\mathcal{B}$ of available data (from $\mathcal{D}_{\mathrm{pre}}$ or $\mathcal{D}_{\mathrm{ft}}$) or \emph{data-agnostic} (ignore $\mathcal{B}$). The superscript $(\cdot)$ indicates that different pruning criteria can be used, as detailed in Sec.~\ref{sec:pruning_strategies}.
%The operator may be \emph{data-independent} (ignore $\mathcal{B}$) or \emph{data-dependent} and use a subset $\mathcal{B}$ of available data (e.g., $\mathcal{D}_{\mathrm{pre}}$ or \revise{$\mathcal{D}_{\mathrm{ft}}$}).
%For online pruning, \revise{$\mathcal{B}\subset\mathcal{D}_{\mathrm{ft}}$} is a subset of pseudo-labeled samples available during operation.

%We define two additional pruning steps that characterize the proposed OnDA pipelines depicted in Fig.~\ref{fig:method_overview}.
On top of the \textsc{Baseline} described in Sec.~\ref{sec:baseline_selflearning}, we identify three ``moments'' at which pruning (and consequent fine-tuning) can be applied, depicted in Fig.~\ref{fig:method_overview} as \textsc{P}, \textsc{O1}, and \textsc{O2}.
%
% Their combinations define the proposed OnDA pipelines.
%
% \revise{We first characterize pruning depending on \textit{when} it is done, i.e., offline or online.}
\textsc{P} is a standard offline pruning that uses pretraining data $\mathcal{D}_{\mathrm{pre}}$ for deciding what to prune (when data-aware) and for fine-tuning the pruned model. %, $\mathcal{A}_{\mathrm{off}}=\mathcal{P}_{\rho_{\mathrm{off}}}^{(\cdot)}(\mathcal{A}_0;\mathcal{D}_{\mathrm{pre}})$, where $\mathcal{A}_0$ is the original architecture of $f_\theta$.
%The second is online pre-adaptation pruning on pseudo-labeled data, $\mathcal{A}_{\mathrm{pre}}=\mathcal{P}_{\rho_{\mathrm{pre}}}^{(\cdot)}(\mathcal{A};\mathcal{B})$ with \revise{$\mathcal{B}\subset\mathcal{D}_{\mathrm{ft}}$}.
In contrast, \textsc{O1} and \textsc{O2} are the two \textit{online} pruning options proposed in this work. In these cases, data-aware criteria and fine-tuning use \textit{pseudo-labeled data} from $\mathcal{D}_{\mathrm{ft}}$. 

\looseness=-1
%Online pruning can be applied either \textit{before} (\revise{\textsc{O1}}) or \textit{after} (\revise{\textsc{O2}}) weights adaptation, i.e., before or after \textsc{B3}.
Online pruning can be applied either \textit{before} (\textsc{O1}) or \textit{after} (\textsc{O2}) weights adaptation (i.e., \textsc{B3}).
\textsc{O2} is the necessary approach for data-agnostic criteria, which decide what to prune not based on input data and corresponding predictions, but rather on weight statistics (e.g., channel norms). Thus, before pruning, such statistics should be updated to reflect the result of in-domain learning, so that the pruning process can retain parts of the DNN that are ``important'' for the context in which it is deployed.  Clearly, in this setting, the network must be fine-tuned again after pruning, to compensate for accuracy drops, hence B3 must be repeated.
In contrast, data-aware criteria, although usually more computationally expensive, can also be applied \textit{prior} to fine-tuning (\textsc{O1}), at the start of the adaptation process, since they prune based on the loss landscape produced by (in-domain) data.

Importantly, both \textsc{O1} and \textsc{O2} are \textit{orthogonal} to \textsc{P}, i.e., they can be applied either directly to unpruned baseline DNNs, or to networks that already underwent an initial offline compression step, using pre-training data ($\mathcal{D}_{\mathrm{pre}}$).

We refer to the complete pipelines obtained by inserting \textsc{O1} or \textsc{O2} at the specified positions into the \textsc{Baseline} sequence of steps (Sec.~\ref{sec:baseline_selflearning}) as \textsc{OnDA-1} and \textsc{OnDA-2} respectively, as depicted in the bottom part of Fig.~\ref{fig:method_overview}.

%\revise{Therefore, we define two pruning pipelines \textsc{OnDA-1} and \textsc{OnDA-2} that extend the \textsc{Baseline} of Sec.~\ref{sec:baseline_selflearning} with \textsc{O1} and \textsc{O2} pruning respectively. Precisely, \textsc{OnDA-1} is the sequence: $\textsc{B1} \rightarrow \textsc{P} (optional) \rightarrow \textsc{B2} \rightarrow \textsc{O1} \rightarrow \textsc{B3}$, whereas \textsc{OnDA-2} corresponds to $\textsc{B1} \rightarrow \textsc{P} (optional) \rightarrow \textsc{B2} \rightarrow \textsc{B3} (1st) \rightarrow \textsc{O2} \rightarrow \textsc{B3} (2nd)$.}

By comparing both approaches, we address the following non-trivial research question: \emph{``Is it preferable to apply in-domain-data-aware pruning prior to fine-tuning, or data-unaware pruning after an initial fine-tuning phase?''}

\subsection{Pruning strategies} \label{sec:pruning_strategies}
As anticipated, we consider structured pruning of the convolutional layer's output channels. We select this compression technique due to its compatibility with commodity infrastructure: the pruned models are essentially dense sub-architectures with fewer channels, hence training and inference speedups and energy savings are easy to achieve.
In contrast, pruning at the level of individual weights or smaller blocks requires dedicated runtimes and/or hardware support to translate into practical efficiency gains~\cite{hoefler2021sparsity}. However, in principle, OnDA can also be applied with unstructured or semi-structured pruning, for hardware platforms supporting them.

Given a layer with output channels $c \in \{1,\dots,C\}$, pruning removes entire activation channels, and the corresponding weights in subsequent layers.
Each layer can have a different number of pruned channels, and we apply pruning criteria globally. That is, all channels in the DNN are scored jointly by an ``importance'' metric, and the least important ones are removed until reaching a target pruning ratio $\rho$, computed as the ratio between the total weights of the pruned model and those of the unpruned DNN. We consider two pruning schemes.

\textbf{Data-agnostic pruning (global L1):}
As data-agnostic method, we apply \textit{magnitude-based} pruning by scoring each channel $c$ with the L1-norm of the weights associated with it as in \cite{li2016pruning}. Concretely, letting $W_c$ denote the kernel slice producing output channel $c$, we define the channel score as:
\begin{equation}
  s^{\text{L1}}_c = \lVert W_c \rVert_1,
\end{equation}
where $\lVert W_c \rVert_1=\sum_{i} |(W_c)_i|$ denotes the sum of absolute values over all scalar entries of $W_c$. We prune the channels with the smallest scores \cite{li2016pruning}. We consider this pruning scheme with the \textsc{OnDA-2} pipeline.

%\subsubsection{Data-aware pruning (HAP)}
\textbf{Data-aware pruning (HAP):}
As data-aware method, we implement Hessian-Aware Pruning (HAP)~\cite{yu2022hessian}.
While magnitude-based criteria ignore how pruning a channel affects the task loss, second-order sensitivity metrics, like HAP, exploit the available data to estimate the local curvature of the loss landscape. Channels whose removal is expected to induce a larger second-order increase in the loss are deemed more important, which is particularly appealing in our low-data setting.
% Let $w_c$ denote the vector of parameters associated with channel $c$.
HAP uses a block-diagonal, trace-based approximation of the Hessian $H_c$ for the group of weights within channel $c$, i.e., $W_c$. Namely, the HAP channel score is proportional to the average Hessian trace scaled by its squared L2-norm and normalized by $n_c$, i.e., the total number of parameters in $W_c$:
\begin{equation}
  s^{\text{HAP}}_c = \Big(\tfrac{\mathrm{Tr}(H_c)}{n_c}\Big)\, \lVert W_c \rVert_2^2.
  \label{eq:hap_score}
\end{equation}
$\mathrm{Tr}(H_c)$ can be efficiently estimated using Hutchinson's method, which computes the application of the Hessian to a random input vector with a computational cost analogous to backpropagating the gradient \cite{yu2022hessian}. We apply HAP pruning in the \textsc{OnDA-1} pipeline and for offline pruning (\textsc{P}).

%To compute data-aware scores $s^{\text{HAP}}_c$ we use a batch $\mathcal{B}$ of either the pre-training data $\mathcal{D}_{\mathrm{pre}}$ or the pseudo-labelled one \revise{$\mathcal{D}_{\mathrm{ft}}$} depending on the specific phase of the considered OnDA pipeline (Sec. \ref{sec:onda_pipelines}).

% \begin{algorithm}[t]
%   \caption{OnDA one-shot online channel pruning before on-device adaptation (P3)}
%   \label{alg:online_prune}
%   \begin{algorithmic}[1]
%     \Require Model parameters $\theta$; target pruning ratio $\rho$; in-field batch $\mathcal{B}$ (first batch of first epoch); criterion $\mathcal{S}\in\{\text{L1},\text{HAP}\}$
%     \Ensure Pruned model $\theta'$
%     \State Compute channel scores $\{s_c\}_{c=1}^{C}$:
%     \If{$\mathcal{S}=\text{L1}$}
%     \State $s_c \leftarrow \lVert W_c \rVert_1$
%     \ElsIf{$\mathcal{S}=\text{HAP}$}
%     \State Estimate $\mathrm{Tr}(H_c)$ on $\mathcal{B}$ via Hutchinson trace and Hessian-vector products
%     \State $s_c \leftarrow (\mathrm{Tr}(H_c)/|w_c|)\,\lVert w_c\rVert_2^2$ \Comment{Eq.~\eqref{eq:hap_score}}
%     \EndIf
%     \State Rank channels by ascending $s_c$ and prune the bottom $\rho$ fraction globally across layers
%     \State Materialize a smaller dense network $\theta'$
%     \State Run on-device adaptation on $\theta'$ using the same gradient-based self-learning scheme as \cite{rusci2024self}
%   \end{algorithmic}
% \end{algorithm}

%
\section{Experimental Results}\label{sec:results}
% \begin{figure*}[!h]
%   \centering
%   \includegraphics[width=\linewidth]{figures/pareto_double_column_stretch.pdf}
%   \vspace{-0.5cm}
%   \caption{Pareto fronts for personalized KWS under different pruning/adaptation pipelines.}
%   \label{fig:pareto}
% \end{figure*}
\begin{figure*}[t]
  \centering
  \includegraphics[width=0.83\linewidth]{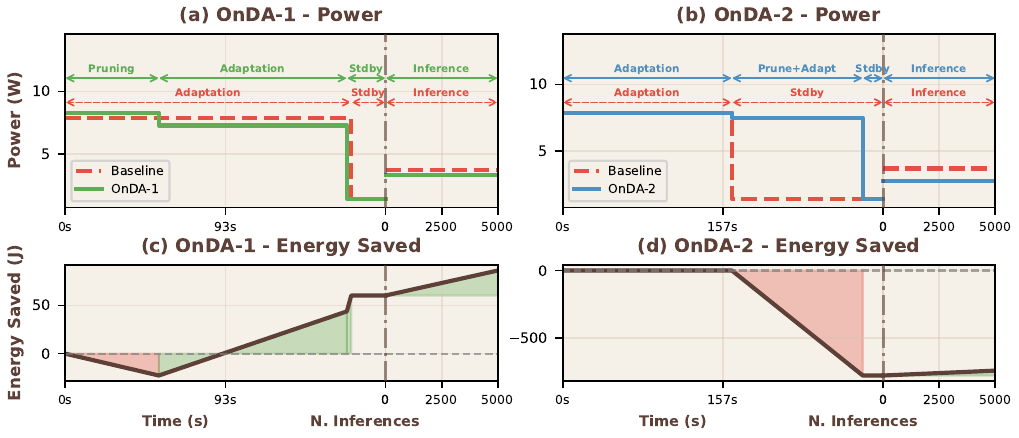}
  \vspace{-0.3cm}
  \caption{On-device latency measurements on Jetson Orin Nano for GPU and CPU deployment.}
  \label{fig:deployment}
  \vspace{-0.5cm}
\end{figure*}
\subsection{Setup}

%We follow the experimental protocol of \cite{rusci2024self} and refer the reader to it for further details. As offline pretraining dataset $\mathcal{D}_{pre}$, we use the MSWC \cite{mazumder2021mswc} large-scale generic keyword dataset. For on-device self-learning and evaluation, we use HeySnips \cite{coucke2019dilated} and HeySnapdragon \cite{qualcommkws} and construct (i) an adaptation set and (ii) a test set as in \cite{rusci2024self}. The test set is used \emph{only} for evaluation before and after self-learning and is composed of (a) positive (keyword) utterances from the same 20 target subjects as \cite{rusci2024self}---using the HeySnips test split for HeySnips subjects and the HeySnapdragon test split for HeySnapdragon subjects---and (b) a shared pool of negative (non-keyword) utterances taken from the HeySnips test split. The adaptation set is used \emph{only} to perform self-learning and model adaptation and contains (a) positive utterances from speakers disjoint from the 20 test subjects (from the corresponding dataset), and (b) additional negative utterances taken from the HeySnips training split; adaptation samples are pseudo-labeled by the pretrained model to form $\tilde{\mathcal{D}}$ and then used for self-learning and model-adaptation.

We follow the experimental protocol of \cite{rusci2024self} and refer the readers to it for further details. We pretrain on the large-scale MSWC \cite{mazumder2021mswc} dataset ($\mathcal{D}_{pre}$) and evaluate baseline and OnDA pipelines on HeySnips \cite{coucke2019dilated} and HeySnapdragon \cite{qualcommkws}, by splitting data into an \textit{adaptation set}, used to build the pseudo-labeled set $\mathcal{D}_{\mathrm{ft}}$ and a disjoint \textit{test set}, used only for evaluation before/after self-learning and model adaptation.
% The test set contains keyword utterances from the same 20 target subjects as \cite{rusci2024self} plus a shared pool of non-keyword utterances from HeySnips; the adaptation set contains keyword utterances from other speakers and additional non-keyword utterances from the HeySnips training split.
We consider the same subjects as \cite{rusci2024self}, and apply each pipeline independently to each subject. We report cross-subject average results. Differently from \cite{rusci2024self}, we repeat all experiments with three different random seeds. When shown, error bars indicate the minimum and maximum across seeds, with the marker denoting the seed-averaged value.

We consider all the models reported in \cite{rusci2024self} as baselines. We apply OnDA to the two largest and best performing architectures, ResNet15 \cite{tang2017resnet} and DS-CNN-L \cite{sainath2015cnn}, and to offline pruned versions of those models with pruning ratios ($\rho$) 25\% and 50\%. We keep the same training/adaptation hyperparameters unless stated otherwise. For each baseline and offline-pruned DNN, we then apply online pruning with ratios 25\%, 50\% and 75\%.

\subsection{OnDA Search Space Exploration} \label{sec:pareto_explo}
Fig.~\ref{fig:pareto} shows the task-performance versus model size trade-off obtained by the baseline and OnDA pipelines. The x-axis reports the model size, while the y-axis reports the accuracy at $\text{FAR}_h=0.5$, i.e., the same task-performance metric of~\cite{rusci2024self}. Different colors denote different architecture families from \cite{rusci2024self}, with the corresponding baseline points connected by a yellow dashed line.
%Different markers indicate different offline pruning ratios, while marker patterns encode the online pipeline: no pattern corresponds to Hessian-based online pruning, a slash pattern corresponds to online $L_1$ pruning, and crosses correspond to the pipeline that performs adaptation, then $L_1$ pruning, and then a second adaptation step.
%Different markers indicate different offline pruning ratios, while marker patterns encode the online pipeline: no pattern and slash pattern correspond to \textsc{OnDA-1} with Hessian-based and $L_1$ pruning respectively, while crosses correspond to \textsc{OnDA-2}.
Different marker shapes indicate different offline pruning ratios, while marker patterns denote the pipeline: no pattern corresponds to \textsc{OnDA-1}, while crosses correspond to \textsc{OnDA-2}.
% Pareto-optimal models are connected by a red dashed line, while
% non-Pareto-optimal models are shown with transparency.
%
Offline pruned DNNs that are not further compressed online are shown with dashed-edge markers.
Lastly, non Pareto-optimal models are shown with transparency.

The results of Fig.~\ref{fig:pareto} show that both OnDA pipelines produce solutions that improve the accuracy versus size trade-off with respect to the baselines, and to DNNs that are only pruned offline, especially in the small size regions of the planes.
We also observe that most of the Pareto-optimal points originate from the ResNet family (orange baseline), suggesting that starting from a larger DNN provides greater flexibility during the subsequent pruning/adaptation optimization.

For HeySnips (left pane), at iso performance with the most accurate baseline (ResNet15), our best OnDA configuration (second orange triangle from the right) achieves $3.33\times$ compression; moreover, the highest iso-performance compression is achieved with respect to the DS-CNN-L baseline, with the second orange square from the left being both $9.63\times$ smaller and more accurate. Considering HeySnapdragon, at iso-performance with ResNet15 baseline, the best compression achieved by our OnDA configurations is $1.7\times$ (second orange circle from the right).

Overall, from the perspective of accuracy versus model size, none of the two OnDA pipelines outperforms the other, with both producing multiple Pareto-optimal solutions. Therefore, the choice between them boils down to their deployment costs, as analysed in the next section.

\begin{table}[t]
  \centering
  \caption{Deployment relative gains on Jetson Orin Nano compute units. Improvements are reported with respect to \cite{rusci2024self}.}
  \label{tab:summary}
  \vspace{-0.2cm}
  \resizebox{0.9\linewidth}{!}{%
    \begin{tabular}{llll}
      \toprule
      Compute unit & OnDA pipeline & Adapt. Lat./En. & Inf. Lat./En. \\
      \midrule
      \textbf{GPU} & \textsc{OnDA-1} & 1.52$\times$/1.64$\times$ & 1.57$\times$/1.77$\times$ \\
      \textbf{GPU} & \textsc{OnDA-2} & 1.29$\times$/1.36$\times$ & 1.91$\times$/2.55$\times$ \\
      \midrule
      \textbf{CPU} & \textsc{OnDA-1} & 1.86$\times$/1.94$\times$ & 1.93$\times$/2.07$\times$ \\
      \textbf{CPU} & \textsc{OnDA-2} & 1.8$\times$/1.87$\times$ & 2.34$\times$/2.53$\times$ \\
      \bottomrule
    \end{tabular}%
  }
\vspace{-0.6cm}
\end{table}

% \begin{table}[t]
%   \centering
%   \caption{Deployment latency and energy on Jetson Orin Nano.}
%   \label{tab:summary}
%   \setlength{\tabcolsep}{3pt}
%   \renewcommand{\arraystretch}{1.15}
%   \scriptsize
%   \begin{tabular}{llccc}
%     \toprule
%     Compute unit & Metric & \textsc{Baseline} & \textsc{OnDA-1-HAP} & \textsc{OnDA-2-L1} \\
%     \midrule
%     \textbf{GPU} & Adapt. Lat. (s) & $166.18$ & $109.26$ & $128.48$ \\
%                 & Adapt. En. (J)  & $1309$ & $794$ & $961$ \\
%                 & Inf. Lat. (ms)   & $3.24$ & $2.06$ & $1.70$ \\
%                 & Inf. En. (mJ)    & $11.95$ & $6.76$ & $4.69$ \\
%     \midrule
%     \textbf{CPU} & Adapt. Lat. (h) & $2.97$ & $1.6$ & $1.65$ \\
%                 & Adapt. En. (kJ)  & $52.38$ & $27.02$ & $29.03$ \\
%                 & Inf. Lat. (ms)   & $69.74$ & $36.12$ & $29.79$ \\
%                 & Inf. En. (mJ)    & $317.18$ & $153.51$ & $125.3$ \\
%     \bottomrule
%   \end{tabular}
% \end{table}

\subsection{Deployment on Jetson Orin Nano}
We deploy OnDA pipelines on the NVIDIA Jetson Orin Nano~\cite{orin_nano}, which integrates both a GPU and a multi-core CPU. We report latency and energy results on both compute units. Power is measured using the on-board current sensor, which provides the System-on-Chip (SoC) current draw.%; we convert this reading into power (W) and integrate over time to estimate energy.

%Based on the results of Sec.~\ref{sec:pareto_explo}, we focus on the two best-performing OnDA pipelines: (i) \revise{\textsc{OnDA-2}} with data-aware (Hessian-based) online pruning, and (ii) \revise{\textsc{OnDA-3}} with data-unaware ($L_1$) pruning. For deployment measurements we select the two highlighted \textcolor{red}{TODO: third and fourth points on the pareto front from right} operating points from the left panel of Fig.~\ref{fig:pareto}, which shows best compression at iso-performance with most accurate baseline.
We analyze deployment metrics for the two circled points from the left panel of Fig.~\ref{fig:pareto}, which are those achieving the best compression at iso-performance compared to the most accurate baseline. Moreover, the two points are very similar in terms of accuracy and model size, but one is produced by \textsc{OnDA-1} and the other by \textsc{OnDA-2}.

% thus allowing us to compare the deployment cost of the two pipelines.

Fig.~\ref{fig:deployment} reports the measured power traces and the resulting energy deltas with respect to the baseline of \cite{rusci2024self} when deploying on the Jetson's GPU. Across all configurations, we keep the same hyperparameters and total training epochs; thus, any change in the duration/power of each fine-tuning/inference step is solely due to architectural differences induced by pruning. In the top panels (a) and (b), the x-axis shows the successive phases of the OnDA pipeline with duration in seconds, followed by a standby phase and then an arbitrary number of inferences $n$ representing normal always-on usage of the adapted model (we report results as a function of $n$ rather than wall-clock time since, in practice, KWS inference can be skipped when the input audio signal energy falls below a threshold). The y-axis reports the measured power consumption in watts. In the bottom panels (c) and (d), the x-axis again reports time/number of inferences, while the y-axis shows the energy difference $\Delta E$ between the baseline and the considered OnDA pipeline (shown in green when positive and in red when negative). Tab.~\ref{tab:summary} summarizes the relative improvements in terms of single inference/fine-tuning step latency and energy for the two pipelines with respect to the \cite{rusci2024self} baseline, both on GPU (visualized in Fig.~\ref{fig:deployment}) and on CPU (not shown for sake of space but exhibiting a similar trend).

An interesting result emerges: although the two pipelines produce similar output models, \textsc{OnDA-1} is drastically faster and more energy efficient when deployed. While HAP pruning is more expensive than L1 in isolation, applying it at the start of the adaptation phase \textit{makes the model smaller} during the subsequent fine-tuning, which in turn becomes faster and draws less power than the baseline. As a result, after a short initial phase with negative $\Delta E$, \textsc{OnDA-1} rapidly becomes more energy efficient than the baseline thanks to a shorter adaptation time and a shorter inference time. This behaviour is not observed for \textsc{OnDA-2}, where the second adaptation step introduces a substantial energy overhead that delays the break-even point to more than $10^5$ inferences. Therefore, we conclude that pre-fine-tuning, data-aware online pruning, is the preferable approach in this scenario.

\section{Conclusions}
We propose OnDA, an online self-learning pipeline for personalized on-device keyword spotting that combines one-shot data-aware pruning with on-device model adaptation. Across HeySnips and HeySnapdragon, OnDA yields Pareto-optimal compressed models that preserve accuracy with $\text{FAR}_{h}=0.5$ while reducing adaptation and inference latency on Jetson Orin Nano by up to 1.52$\times$/1.57$\times$.
%
% \input{sections/6_generative_ai_disclosure}
%\newpage
\bibliographystyle{IEEEtran}
\bibliography{mybib}

\end{document}